\documentclass{anstrans}
\title{Algorithms for TRISO Fuel Identification Based on X-ray CT Validated on Tungsten-Carbide Compacts}
\author{Ming Fang, Angela Di Fulvio}

\institute{
Department of Nuclear, Plasma, and Radiological Engineering, University of Illinois at Urbana-Champaign, \\104 S. Wright St., Urbana, IL 61801, mingf2@illinois.edu, difulvio@illinois.edu
}

\usepackage{graphicx} 
\usepackage{booktabs} 
\usepackage{microtype} 

\usepackage{caption}
\usepackage{subcaption}
\usepackage{hyperref}
\usepackage[dvipsnames]{xcolor}
\usepackage[final]{changes}

\begin{document}
\section{Introduction}
Tristructural-isotropic (TRISO) fuel is one of the most mature fuel types for candidate advanced reactor types under development. TRISO-fueled units consist of thousands fissile kernels of 350-600 µm diameter surrounded by graphite and carbonized resin in the form of a billiard ball-size pebble or a compact cylinder.\deleted{ Each uranium-bearing kernel of 350-600 µm diameter is surrounded by coating layers for retaining gaseous fission products. }This unique design makes TRISO fuel structurally more resistant to neutron irradiation, corrosion, oxidation and high temperatures than traditional reactor fuels. \replaced{TRISO-fuel pebbles flow continuously through the reactor core and can be reinserted into the reactor several times until a target burnup is reached. The capability of identifying individual fuel pebble would allow us to calculate the fuel residence time in the core and validate pebble flow computational models, prevent excessive burnup accumulation or premature fuel discharge, and maintain accountability of special nuclear materials during fuel circulation. }{Non-destructive techniques to identify individual pebbles are required to help validate pebble flow computational models and prevent excessive burnup accumulation or premature fuel discharge.}

The three dimensional (3D) distribution of TRISO particles in the pebble makes each fuel unit inherently unique. X-ray computed tomography (CT) and associated image processing algorithms can be used to image the unique structure of the fuel matrix and extract it as an effective fingerprint to identify a single pebble. In our previous work~\cite{Ming2021}, we have developed an accelerated Monte Carlo algorithm to produce X-ray CT scans of TRISO fuel pebbles and validated it using MCNP6.2. In this work, we have developed a 3D image reconstruction and segmentation algorithm to accurately segment TRISO particles and extract the unique 3D distribution. The main challenge of the identification is that the irradiated pebble could have undergone an arbitrary rotation in the reactor core, and the extracted 3D particle distribution might be completely different from the previous one. We have developed a rotation-invariant and noise-robust identification algorithm that allows us to identify the pebble and retrieve the pebble ID in the presence of rotations and noises. We also report the results of 200kV X-ray CT image reconstruction of a mock-up fuel sample consisting of tungsten-carbide (WC) kernels in a lucite matrix.\added{ The 3D distribution of TRISO particles along with other signatures such as $^{235}$U enrichment and burnup level extracted through neutron multiplicity counting, would enable accurate fuel identification in a reasonable amount of time.}


\section{Methods}

\subsection{Generation of X-ray Images}
\begin{figure}[!htbp]
    \captionsetup{font=footnotesize}
    \centering
	\includegraphics[width=.5\linewidth]{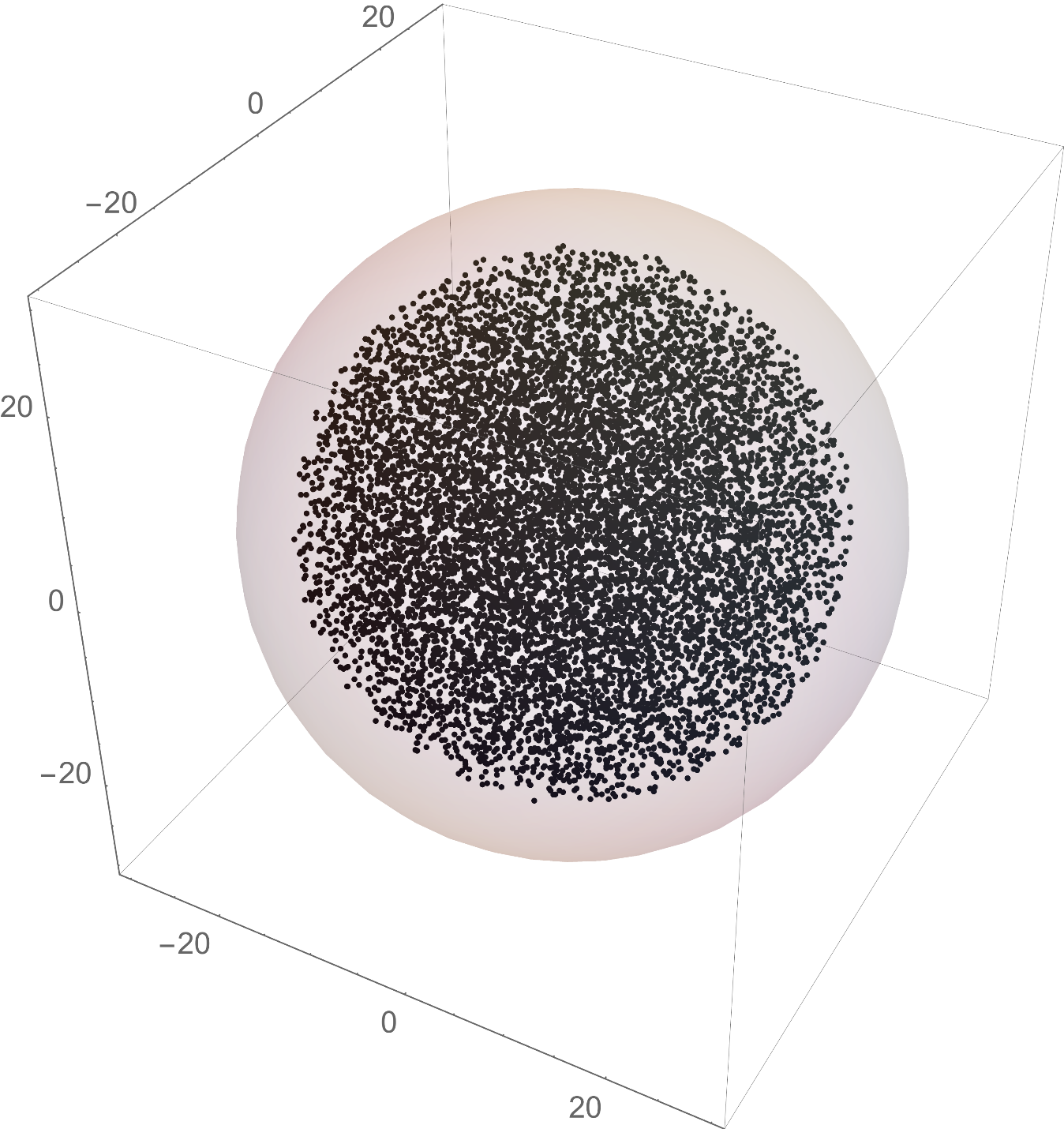}
	\caption{3D visualization of the pebble model (unit: mm). Black dots represent the TRISO particles.}
	\label{fig:pebble3d}
\end{figure}

We first created a 3D model of the pebble shown in Fig.~\ref{fig:pebble3d}. The pebble is a 3-cm radius sphere filled with TRISO particles surrounded by graphite. Around 10k randomly-distributed TRISO particles were created inside a 2.5-cm radius sphere, leaving a 5-mm margin at the boundary. Table~\ref{table:exp_uco} shows the material composition of the UCO kernel in the TRISO particle. The material of each TRISO particle was homogenized to simplify the computation while keeping a realistic composition. 
\begin{table}[!htbp]
	\centering
    \captionsetup{font=footnotesize}
	\caption{The mole fraction and density of each component in UCO kernel.}\label{table:exp_uco}
    \begin{tabular}{l|ccc}
                      & \textbf{$\text{UO}_2$}  & \textbf{UC} & \textbf{$\text{UC}_2$}   \\ \hline
    \textbf{Mole fraction}~\cite{helmreich2017year}      & 0.69 & 0.03  & 0.28  \\ \hline
    \textbf{Density (g/cm$^3$)}~\cite{holleck2013u} & 10.8 & 11.69 & 13.60
    \end{tabular}
\end{table}

We have implemented an accelerated Mote Carlo algorithm to generate synthetic X-ray images of fuel pebbles, accounting for the attenuation and scattering effects~\cite{Ming2021}. The simulated X-ray tube voltage was 250~kVp, which is typically employed in industrial X-ray computed tomography and yields a Bremsstrahlung spectrum of 87~keV average energy with characteristic X-ray peaks. A parallel X-ray beam was generated to irradiate the pebble sphere. The X-ray energy spectrum was simulated by SpekCalc~\cite{poludniowski2009spekcalc}. The imaging area was 600 × 600 pixels, each pixel of 100~$\mu$m × 100~$\mu$m. We first considered the contribution of primary X-rays that penetrate the pebble sphere without any attenuation, based on Beer-Lambert law. Then we tracked the photons' travel in the sphere and calculated the contributions of all scattered photons using the forced-detection technique~\cite{fang2021quantitative}. We ran the simulation for $10^7$ source photons to achieve good statistics. The two contributions were summed to form an X-ray image. We generated 360 X-ray images with an angle step of $1^\circ$.

\subsection{Image Reconstruction}
We have implemented a 3D image reconstruction algorithm based on the filtered back-projection (FBP). Let $i,j,k$ be the index of scanning angle, row, and column of the X-ray images, respectively. We first restructure the X-ray images based on index $j$, i.e., put the X-ray scans of the same slice of the pebble at different angles together to form a 2D sinogram of each slice. We then applied the FBP algorithm to reconstruct a 2D image of each slice and stacked all 2D images together to form a 3D image. 
\subsection{Image Segmentation}
Given the reconstructed image, we aimed to identify the TRISO particles inside the pebble and determine their locations through image segmentation. Thresholding was used as the automated segmentation method for its simplicity and robustness. 
We used Otsu's method to determine an intensity threshold that minimizes the intra-class variance, considering one intensity class as the TRISO particles' intensity levels with other class the surrounding graphite~\cite{otsu1979threshold}. We then created a 2D mask by labeling the pixels with values above the threshold as 1 and all the other pixels as 0. We used the Dice's coefficient as a measure of segmentation performance~\cite{dice1945measures}. The Dice's coefficient quantifies the similarity of two samples:
\begin{equation}
    \text{D} = 2\times \frac{\text{overlap of two images}}{\text{sum}(\text{image } A) + \text{sum}(\text{image } B)},
    \label{eq:Dice}
\end{equation}
where image $A$ is the segmented image and image $B$ is the ground truth. A higher Dice's coefficient means better segmentation.

All the 2D masks were stacked together to form a 3D mask. We identified the interconnected regions on the 3d mask as the particles and measured their centroids and radii using the Scipy package~\cite{scikit-image}. We calculated the identification percentage, which is the ratio between the number of identified particles and true number of particles.

\subsection{Pebble Identification}
Given the 3D spatial distribution of the TRISO particles (point cloud), the task is to compare it with the existing fuel pebbles in the library and extract the pebble ID. The main challenge of identification is that the inspected pebble could have undergone an arbitrary rotation, and the resulting point cloud is completely different from the ground truth in the library. To solve this, we searched over all possible rotations and determined a rotation matrix that minimizes the difference between the rotated point cloud and ground truth. We used the global iterative closest point (Go-ICP) algorithm to find the optimal rotation matrix~\cite{yang2015go}. We denote the point cloud of the inspected pebble as data: $X=\{x_i\in \mathcal{R}^3\}_{i=1}^N$ and the point cloud of the ground truth as model: $Y=\{y_i\in \mathcal{R}^3\}_{i=1}^M$. We aimed to minimize the sum of squared distances between all pairs of closest points in data and model:
    \begin{equation}
        E(R) = \sum_{i=1}^N \|Rx_i-y_{j^*}\|^2,
    \end{equation}
where $R$ is a rotation matrix, and $y_{j^*}$ is the closest point to $Rx_i$:
    \begin{equation}
        j^* = \underset{j\in {1,\cdots M}}{\arg\min}\|Rx_i -y_j\|. 
    \end{equation}
Once the optimal rotation matrix is found, we can calculate the minimized error $E(R)$. If the minimized error is below a threshold, we conclude that the two pint clouds match and the pebble ID is found. If the minimized error is above the threshold or the optimization does not converge within the maximum number of searches, we conclude that data and model do not match. 
   
We created a library of 100 fuel pebbles and applied a random rotation to each pebble to examine the performance of our identification algorithm. We then used the algorithm to find the ID of each rotated pebble by comparing it with each pebble in the library. To speed up the comparison, we consider only points that are 5-6~mm away from the pebble center. To examine the robustness of our identification algorithm against noise, we added Gaussian noise to the data and calculated the identification ratio as a function of the noise level.
\subsection{Scan of Mockup Samples}
\begin{figure}[!htbp]
    \captionsetup{font=footnotesize}
    \centering
     \includegraphics[width=\linewidth]{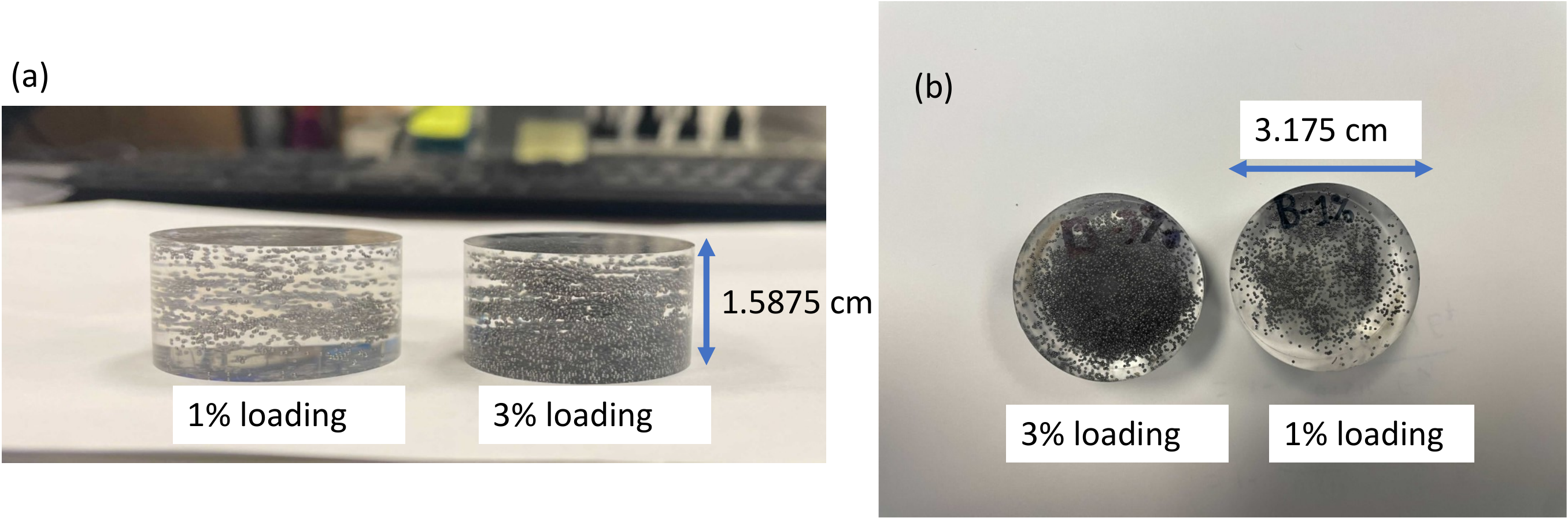}
	\caption{Side view (a) and top view (b) of two WC samples.}
  \label{fig:wc_samples}
\end{figure}

We performed cone-beam X-ray CT scans of two tungsten-carbide (WC) samples using a North Star X5000 industrial CT scanner. The WC samples were made by mixing tungsten carbide particles of 500-$\mu$m diameter with Lucite thermoplastic metallographic mounting material (LECO 811-132). WC has a deinsty of 15.63~g/cc and Lucite has a density of 1.18~g/cc. Fig.~\ref{fig:wc_samples} shows two WC samples with 1\% WC volume loading fraction and 3\% WC volume loading fraction, which have comparable attenuation as a 17\% and 50\% particle-loading fuel compact, respectively.

The X-ray tube voltage was set to 200 kVp and the current was 36 $\mu$A. The detector panel has 1536x1920 pixels, with a pixel pitch of 127 $\mu$m. The angle increment was ${0.1}^{\circ}$ and totally 3600 projection were acquired. The 3D image reconstruction was performed using the reconstruction software by North Star Inc. We then applied our segmentation algorithm to extract the positions of WC particles. 


\section{Results}
\begin{figure}[!htbp] 
    \captionsetup{font=footnotesize}
    \centering
     \includegraphics[width=.65\linewidth]{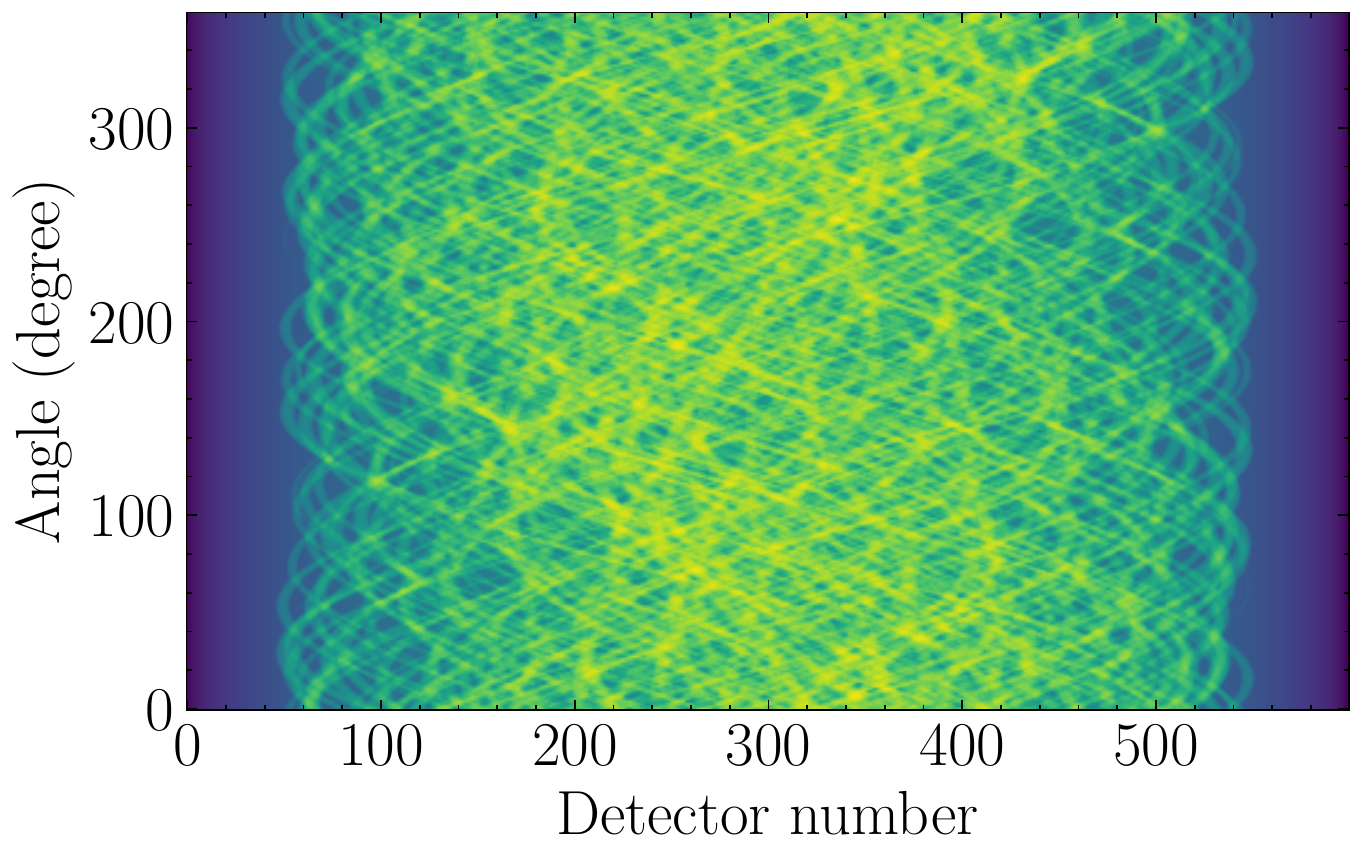}
	\caption{The simulated sinogram of a single slice of the fuel pebble.}
  \label{fig:sino}
\end{figure} 
\subsection{Image Reconstruction}
\begin{figure}[!htbp]
    \captionsetup{font=footnotesize}
    \centering
      \includegraphics[width=.85\linewidth]{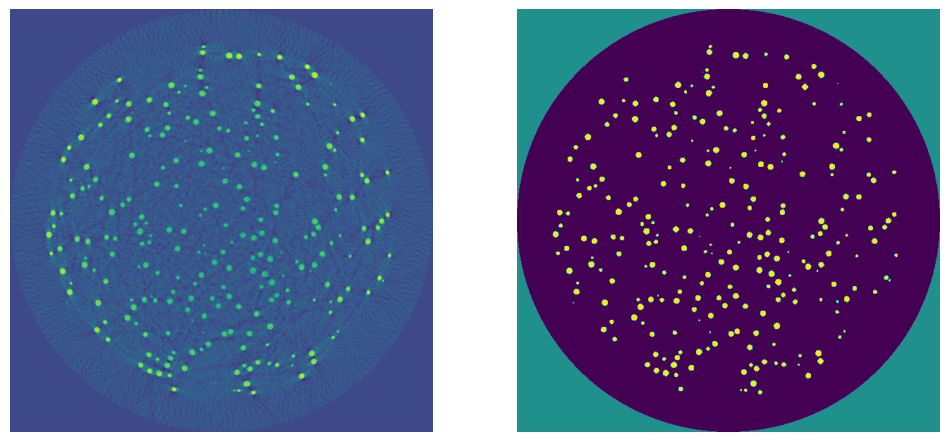}
      \caption{Image reconstruction based on Fig.~\ref{fig:sino}. (Left) Image reconstructed from the sinogram in Fig.~\ref{fig:sino}. (Right) Ground truth image of the same slice.}
      \label{fig:reco}
      \vspace{-2em}
\end{figure}

Fig.~\ref{fig:sino} shows the 2D sinogram of a slice of the pebble. We applied the FBP algorithm to reconstruct an image of the slice. Fig.~\ref{fig:reco} shows the reconstructed image and the ground truth. We observe a good similarity between the two images. 

\subsection{Image Segmentation}

\begin{figure}[!htbp]
    \captionsetup{font=footnotesize}
    \centering
    \includegraphics[width=.85\linewidth]{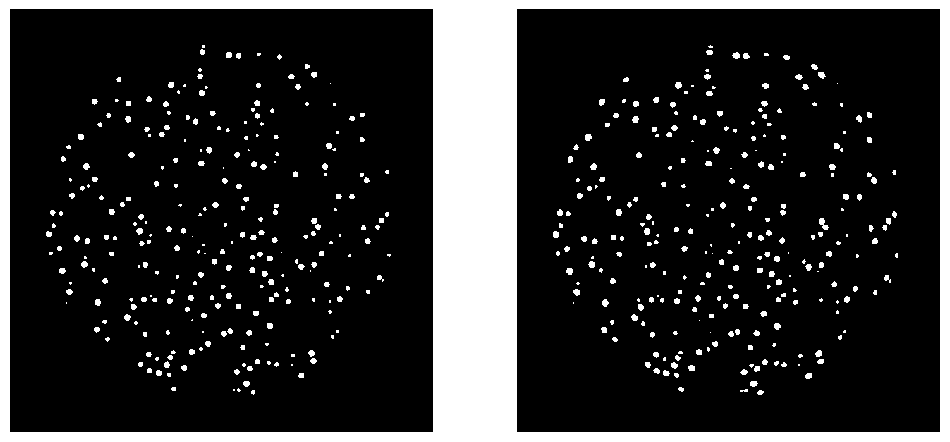}
    \caption{Image segmentation based on Fig.~\ref{fig:reco}. (Left) Segmented image (mask). (Right) Ground truth.}
    \label{fig:imgseg}
\end{figure}

Fig.~\ref{fig:imgseg} shows an example of 2D image segmentation. The segmentation algorithm correctly labels pixels within the TRISO paricle as 1 despite the small areas of some bright regions.

\begin{figure}[!htbp]
    \captionsetup{font=footnotesize}
    \centering
    \includegraphics[width=.5\linewidth]{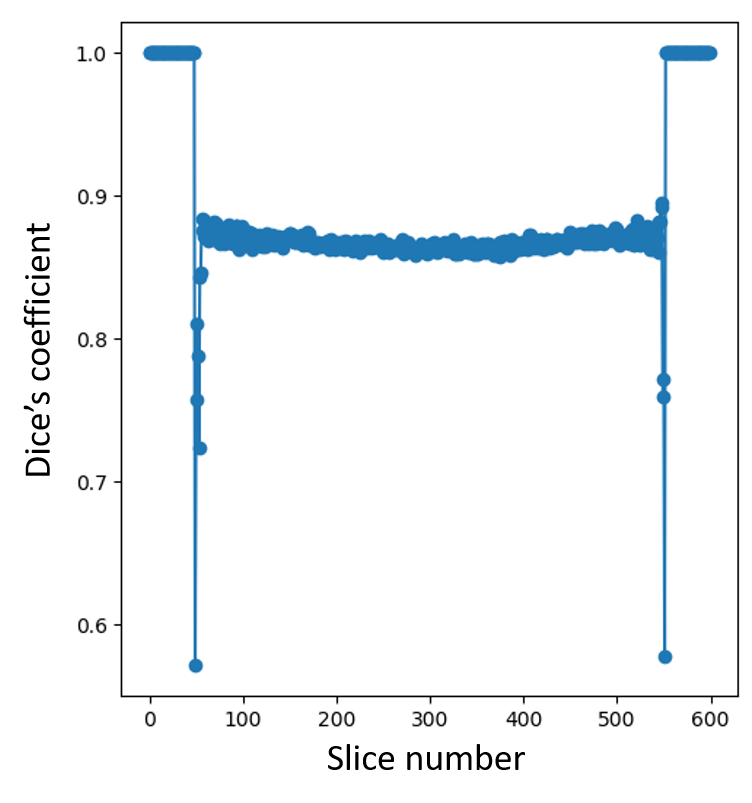}
    \caption{Dice's coefficient of each segmentation. A nearly uniform Dice score of 87\% was achieved for all slices within the fuel zone. Dice score of slices below 50 and above 550 should be neglected since no TRISO particles are present in the pebble periphery. }
    \label{fig:dice} 
\end{figure}

We calculated the Dice's coefficient for each segmented image, shown in Fig.~\ref{fig:dice}. For most slices, we obtained a Dice's coefficient of 87\%. At slice 50 and slice 550, the Dice's coefficient is low because the slicing plane is tangent to the fuel zone, resulting in few non-zero pixels. {The Dice score of slices below 50 and above 550 should be neglected since no TRISO particle are present in the pebble periphery.}

\begin{figure}[!htbp]
    \captionsetup{font=footnotesize}
    \centering
    \includegraphics[width=.5\linewidth]{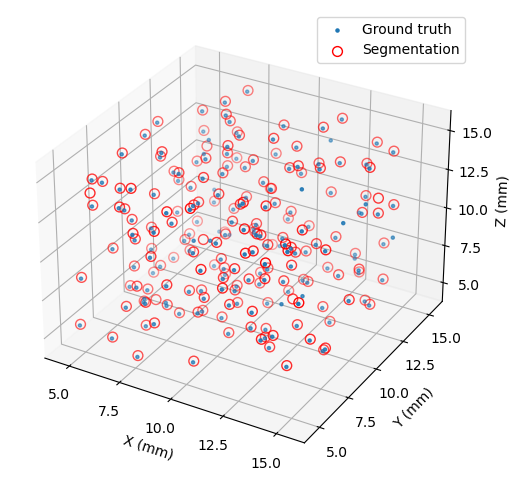}
    \caption{Comparison of the reconstructed TRISO particles and the ground truth. For visualization purposes, particles with $x,y,z$ coordinates between 5~mm ans 15~mm are shown and the radius of the red circle does not stand for the true pebble radius. }
    \label{fig:3drecon}
\end{figure}

Fig.~\ref{fig:3drecon} shows the comparison between the reconstructed TRISO particles (red circles) and the ground truth (blue dots). The segmentation successfully identifies most of TRISO particles in the fuel pebble, with a few outliers due to image reconstruction artifacts. Overall, we correctly reconstructed the positions of 9664 TRISO particles in a pebble containing 10254 particles, leading to an identification percentage of 94\%. It took approximately 30 seconds to segment the 3d image and extract the positions of TRISO particles.
\subsection{Pebble Identification}
\begin{figure}[!htbp]
    \captionsetup{font=footnotesize}
    \centering
    \includegraphics[width=\linewidth]{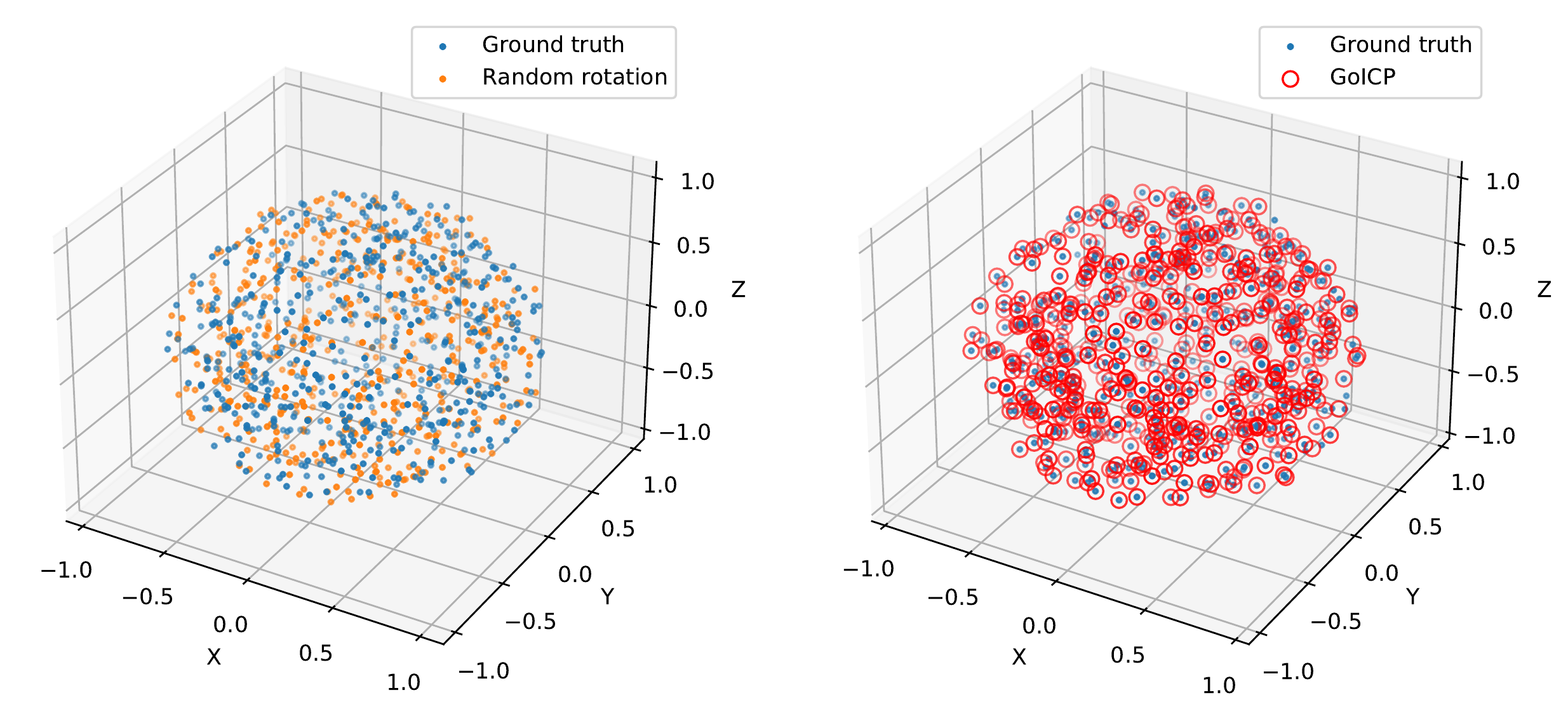}
    \caption{Alignment of two point clouds using Go-ICP. (Left) The model (ground truth, blue dot) and data (yellow dot) taken from the true pebble and reconstructed pebble (randomly rotated), respectively. (Right) The data after applying an inverse rotation calculated by Go-ICP, which agrees well with the ground truth. The X, Y, Z axes are normalized.}
    \label{fig:goicp}
\end{figure}

In Fig.~\ref{fig:goicp}, we show the alignment of a randomly rotated pebble with the corresponding ground truth, with the rotation matrix calculated by Go-ICP. The algorithm is able to find the correct rotation that the reconstructed pebble has gone through, despite the outliers and noise artifacts introduced during the image reconstruction and segmentation.

\begin{figure}[!htbp]
    \captionsetup{font=footnotesize}
\begin{subfigure}[t]{0.45\linewidth}
    \centering
    \includegraphics[width=\linewidth]{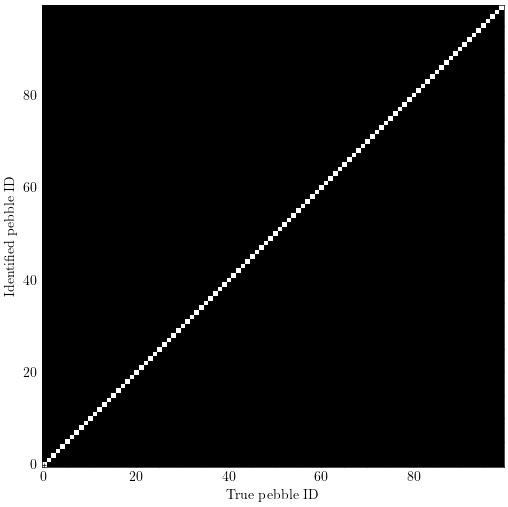}
    \caption{Identification of 100 randomly rotated pebbles. Each row is obtained by comparing a rotated pebble with all 100 pebbles in the library. The white pixel stands for a match between the model and data.}
    \label{fig:pebbleid}
\end{subfigure}\hfil
\begin{subfigure}[t]{0.45\linewidth}
    \centering
    \includegraphics[width=\linewidth]{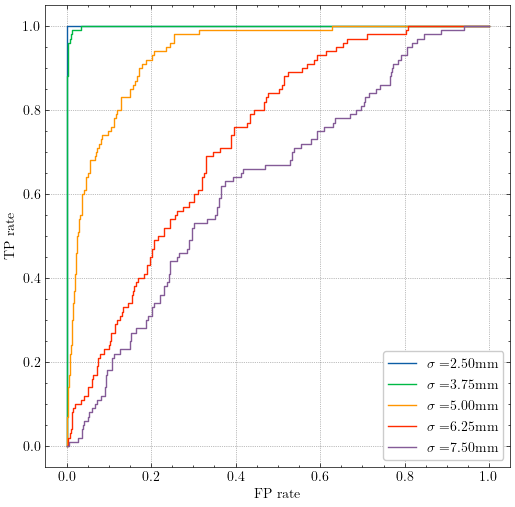}
    \caption{ROC curves at different noise levels. $\sigma$ is the standard deviation of the Gaussian noise applied to the positions of TRISO particles. The ROC curve is obtained by varying the identification thresholds.}
    \label{fig:pebbleidROC}
\end{subfigure}
\end{figure}

The identification result of 100 randomly-rotated pebbles is shown in Fig.~\ref{fig:pebbleid}. We successfully identified all the pebbles regardless of the rotation angle and axis. It took approximately 30 seconds to identify a single rotated pebble. We applied a Gaussian noise to the points and obtained the receiver operating characteristic (ROC) curves at various noise levels shown in Fig.~\ref{fig:pebbleidROC}. The algorithm performs well in the presence of high noise level (standard deviation up to 3.75~mm, which is 7.5x particle's diameter), showing that the identification algorithm is robust against noise.

\subsection{Scan of Mockup Samples}
\begin{figure}[!htbp]
    \captionsetup{font=footnotesize}
    \begin{subfigure}[t]{0.45\linewidth}
        \centering
        \includegraphics[width=\linewidth]{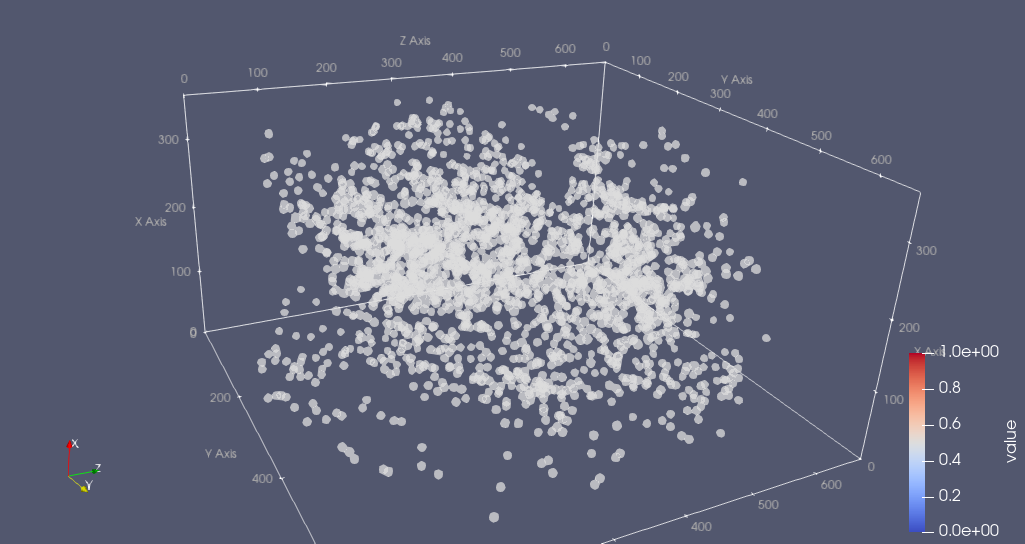}
        \caption{ }
        \label{fig:1p_recon}
    \end{subfigure}\hfil
    \begin{subfigure}[t]{0.45\linewidth}
        \centering
        \includegraphics[width=\linewidth]{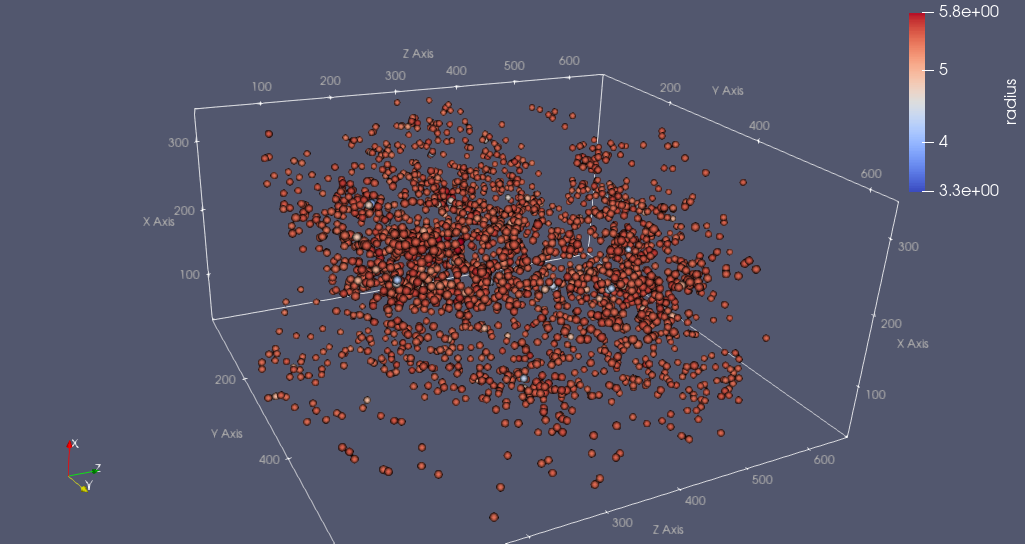}
        \caption{ }
        \label{fig:1p_segmt}
    \end{subfigure}
    \caption{Reconstructed image (left) and segmented particles (right) of 1\% WC loading sample.}
\end{figure}

\begin{figure}[!htbp]
    \captionsetup{font=footnotesize}
    \begin{subfigure}[t]{0.45\linewidth}
        \centering
        \includegraphics[width=\linewidth]{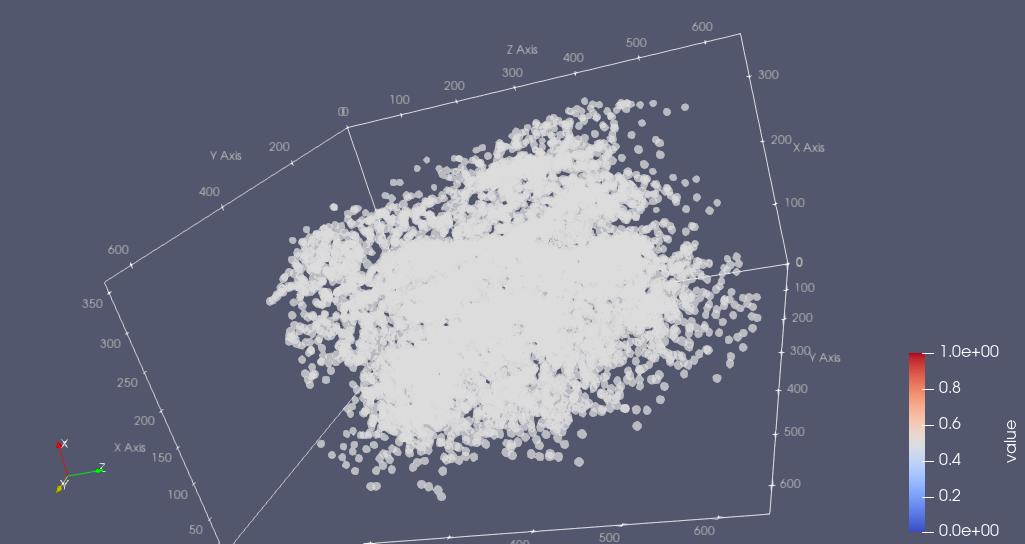}
        \caption{ }
        \label{fig:3p_recon}
    \end{subfigure}\hfil
    \begin{subfigure}[t]{0.45\linewidth}
        \centering
        \includegraphics[width=\linewidth]{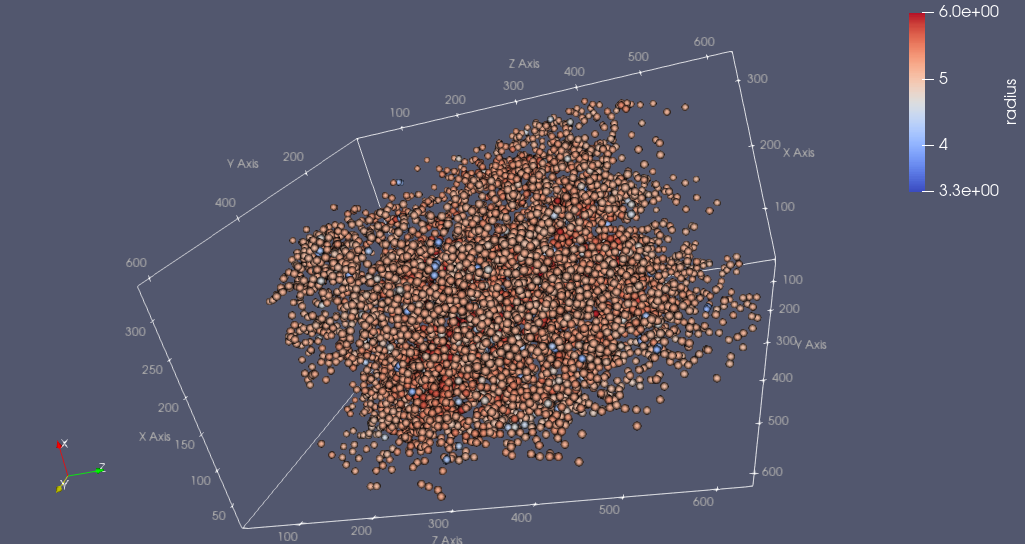}
        \caption{ }
        \label{fig:3p_segmt}
    \end{subfigure}
    \caption{Reconstructed image (left) and segmented particles (right) of 3\% WC loading sample.}
\end{figure}

Fig.~\ref{fig:1p_recon} and Fig.~\ref{fig:3p_recon} show the reconstructed image of 1\% and 3\% WC loading samples. Fig.~\ref{fig:1p_segmt} and Fig.~\ref{fig:3p_segmt} show the identified particles\footnote{Videos showing the full 3D view can be accessed at \url{https://uofi.box.com/v/200kVp-3dvideo}}. We observed a good agreement between the two, showing the segmentation algorithm's capability of handling real images. The number of identified particles is 2231 for 1\% WC loading sample and 6651 for 3\%, approximately 15\% higher than the true number. The over-segmentation is due to the noises and reconstruction artifacts and can be mitigated by pre-procssing the image before segmentation, such as denoising and smoothing.

\section{Conclusions}
In this work, we have developed a set of algorithms for X-ray CT based TRISO fuel identification, including 3D image reconstruction and segmentation algorithms to accurately segment TRISO particles and extract their unique 3D distribution, and a point-cloud registration-based identification algorithm that allows to identify the pebble and retrieve the pebble ID in the presence of noises and arbitrary rotations. We performed 10,000 computations using the identification algorithm and achieved an identification ratio of 100\%. We added Gaussian noises to the particle distribution of the rotated pebble to examine the noise-robustness of the identification algorithm and no performance degradation was observed with standard deviation up to 7.5x particle diameter. It took approximately 30 seconds to segment the 3D image and 30 seconds to identify a single rotated pebble with a single core computer running a single thread. We applied the segmentation algorithm to 200kVp X-ray CT scans of mockup fuel samples and achieved a segmentation ratio of 85\%.
\section{Acknowledgments}
This work was funded in part by the STTR-DOE grant DE-SC0020733 and by the Nuclear Regulatory Commission Faculty Development Grant number 31310019M0011. We also would like to thank Mr. Kevin Hour (BWX Technologies Inc.) for providing the WC samples.

\bibliographystyle{ans}
\bibliography{bibliography}

\begin{thebibliography}{1}
\newcommand{\enquote}[1]{``#1''}

\bibitem{Ming2021}
\MakeUppercase{M.~Fang} and \MakeUppercase{{Di Fulvio, Angela}},
  \enquote{Multi-Mode Imaging for TRISO-fueled Pebble Identification,}
  \emph{ANS Student Conference 2021} (2021).

\bibitem{helmreich2017year}
\MakeUppercase{G.~W. Helmreich}, \MakeUppercase{J.~D. Hunn},
  \MakeUppercase{J.~W. McMurray}, \MakeUppercase{R.~D. Hunt},
  \MakeUppercase{B.~C. Jolly}, \MakeUppercase{M.~P. Trammell},
  \MakeUppercase{D.~R. Brown}, \MakeUppercase{B.~J. Blamer},
  \MakeUppercase{T.~J. Reif}, and \MakeUppercase{H.~T. Kim}, \enquote{Year One
  Summary of X-energy Pebble Fuel Development at ORNL,} Tech. rep., Oak Ridge
  National Lab.(ORNL), Oak Ridge, TN (United States) (2017).

\bibitem{holleck2013u}
\MakeUppercase{H.~Holleck} and \MakeUppercase{H.~Kleykamp}, \emph{U Uranium:
  Uranium Carbides}, Springer Science \& Business Media (2013).

\bibitem{poludniowski2009spekcalc}
\MakeUppercase{G.~Poludniowski}, \MakeUppercase{G.~Landry},
  \MakeUppercase{F.~Deblois}, \MakeUppercase{P.~Evans}, and
  \MakeUppercase{F.~Verhaegen}, \enquote{SpekCalc: a program to calculate
  photon spectra from tungsten anode x-ray tubes,} \emph{Physics in Medicine \&
  Biology}, \textbf{54}, \emph{19}, N433 (2009).

\bibitem{fang2021quantitative}
\MakeUppercase{M.~Fang}, \MakeUppercase{Y.~Altmann},
  \MakeUppercase{D.~Della~Latta}, \MakeUppercase{M.~Salvatori}, and
  \MakeUppercase{A.~Di~Fulvio}, \enquote{Quantitative imaging and automated
  fuel pin identification for passive gamma emission tomography,}
  \emph{Scientific Reports}, \textbf{11}, \emph{1}, 1--11 (2021).

\bibitem{otsu1979threshold}
\MakeUppercase{N.~Otsu}, \enquote{A threshold selection method from gray-level
  histograms,} \emph{IEEE transactions on systems, man, and cybernetics},
  \textbf{9}, \emph{1}, 62--66 (1979).

\bibitem{dice1945measures}
\MakeUppercase{L.~R. Dice}, \enquote{Measures of the amount of ecologic
  association between species,} \emph{Ecology}, \textbf{26}, \emph{3}, 297--302
  (1945).

\bibitem{scikit-image}
\MakeUppercase{S.~van~der Walt}, \MakeUppercase{J.~L. {S}ch\"onberger},
  \MakeUppercase{J.~{Nunez-Iglesias}}, \MakeUppercase{F.~{B}oulogne},
  \MakeUppercase{J.~D. {W}arner}, \MakeUppercase{N.~{Y}ager},
  \MakeUppercase{E.~{G}ouillart}, \MakeUppercase{T.~{Y}u}, and
  \MakeUppercase{the scikit-image contributors}, \enquote{scikit-image: image
  processing in {P}ython,} \emph{PeerJ}, \textbf{2}, e453 (6 2014).

\bibitem{yang2015go}
\MakeUppercase{J.~Yang}, \MakeUppercase{H.~Li}, \MakeUppercase{D.~Campbell},
  and \MakeUppercase{Y.~Jia}, \enquote{Go-ICP: A globally optimal solution to
  3D ICP point-set registration,} \emph{IEEE transactions on pattern analysis
  and machine intelligence}, \textbf{38}, \emph{11}, 2241--2254 (2015).

\end{thebibliography}
\end{document}